\documentclass{article} 

\usepackage{amsmath,amssymb,amsthm,latexsym} 

\usepackage{stmaryrd,wasysym,upgreek,mathrsfs,dsfont} 
\usepackage[english]{babel} 
\usepackage{graphicx,color} 

\newcommand{\bbbone}{{\mathds{1}}}
\newtheorem{definition}{Definition}[section]
\newtheorem{theorem}{Theorem}[section]
\newcommand{\be}{\begin{equation}}
\newcommand{\ee}{\end{equation}}
\newcommand{\bea}{\begin{eqnarray}}
\newcommand{\eea}{\end{eqnarray}}

\newcommand{\cE}{{\cal{E}}}

\newcommand{\cP}{{\cal{P}}}
\newcommand{\cF}{{\cal{F}}}
\newcommand{\cS}{{\cal{S}}}

\begin{document} 
 \title{Tree Quantum Field Theory} 
\author{R. Gurau$^{1}$, J. Magnen$^{2}$, V. Rivasseau$^{1}$\\
1) Laboratoire de Physique Th\'eorique, CNRS UMR 8627,\\ 
Universit\'e Paris XI,  F-91405 Orsay Cedex, France\\
2) Centre de Physique Th\'eorique, CNRS UMR 7644,\\
Ecole Polytechnique F-91128 Palaiseau Cedex, France}

\maketitle 
\begin{abstract} 
We propose a new formalism for quantum field theory (QFT) 
which is neither based on functional integrals, 
nor on Feynman graphs, but on marked trees. This formalism is constructive, i.e. it 
computes correlation functions through convergent
rather than divergent expansions. It applies both
to Fermionic and Bosonic theories. It is compatible
with the renormalization group, and it allows to define non-perturbatively
{\it differential} renormalization group equations. 
It accommodates any general stable polynomial Lagrangian.
It can equally well treat noncommutative models or matrix models 
such as the Grosse-Wulkenhaar model. Perhaps most importantly
it removes the space-time background from its central place in QFT,
paving the way for a nonperturbative definition of field
theory in noninteger dimension.
\end{abstract} 

\section{Introduction} 

Feynman invented the two pillars of quantum field theory (or QFT): functional integrals
and Feynman graphs. However  none of them is fully satisfactory. Indeed
QFT (and in particular its soul, renormalization theory) requires to compute
connected functions. Functional integrals give rise to singular limits such as $0/0$ for such 
connected functions in infinite volume. Feynman graphs apparently solved this problem because
connected functions are expressed as the sum of connected graphs. However the price is too heavy:
perturbation theory based on Feynman graphs always diverge because there are 
too many graphs at large order. Since any alternate divergent series can be cut into pieces
and rearranged to converge to any number we want, ordinary perturbative QFT 
does not define anything at a fundamental level. The vast majority of quantum field theorists 
(with the exception of the small tribe of constructive field theorists) 
essentially pulls this fundamental problem under the rug. However the 
tremendous achievements of standard quantum field theory should not be denied either.
Functional integrals can be simulated, for instance for QCD through Monte Carlo numerical experiments.
Perturbative computations in QED allow incredibly accurate comparisons of theory and experiments
thanks to the smallness of the fine structure constant. But there is no proof (and even no expectation) that these successful computations should converge better and better with more and more computing power. They do not therefore by themselves constitute a true theory in any fundamental sense.
We are also well aware about the very interesting axiomatic or algebraic approaches to QFT.
However these approaches did not lead up to now to the construction of interacting quantum field theories.  Only constructive field theory in the 70's and 80's
succeeded in defining rigorously some interacting QFT's
but in dimensions less than four. But we must admit that the ugly
constructive tools (truncated functional integrals, cluster and Mayer expansions, "large/small"
field expansions) which were neither canonical nor optimal, largely prevented 
the spread of that approach beyond a small circle of {\it aficionados}.

Perhaps axiomatization of QFT might have been premature.
Indeed new field theories constantly arise in an extended sense.
Condensed matter is clearly better understood in a field theoretic formalism, although
that formalism is not relativistic and has finite density. More recently
noncommutative QFTs has been shown renormalizable. They show amazing similarities and subtle
differences with ordinary QFT.

Ultimately we think that combinatorics is the right approach to QFT and that 
a QFT should be thought of as the generating functional of a
certain weighted species in the sense of \cite{BLL}.

In this paper we perform a step in this direction: we 
show how to base quantum field theory
on trees, which lie at the right middle point between
functional integrals and Feynman graphs so that they 
share the advantages of both, but none of their problems.

The core of our proposal is to distinguish among model independent and 
model dependent aspects of QFT.
There are three model-independent ingredients: a universal vector space 
algebraically spanned by all marked trees,
a universal "canonical Hamiltonian operator" which essentially
glues a new subtree at the mark on the tree, and the canonical forest formula
of \cite{BK,AR1}, which is promoted to a central tool of quantum 
field theory.

A particular (Euclidean) quantum field theory model is a particular positive scalar 
product\footnote{To treat Minkowski signature we need to extend 
our definition so as to allow nondegenerate but not 
necessarily positive scalar products. This will not be studied here.}
on the universal vector space. That scalar product is simply obtained
by applying the canonical forest formula to the ordinary perturbative expansion 
of the considered QFT model.
The canonical formula itself is model-independent. What that magic formula does
is conceptually not difficult to understand. It just classifies 
Feynman amplitudes differently, 
by breaking these amplitudes into pieces and putting these pieces into boxes 
labeled by trees. The important point is that it does this in a canonical,
``democratic'', positivity preserving way.

Model-dependent details such as space-time dimension, 
interactions and propagators are
therefore no longer considered fundamental.
They just enter the definition of the matrix elements 
of this scalar product. These matrix elements are just finite sums 
of finite dimensional Feynman integrals. It is just
the packaging of perturbation theory which is redone 
in a better way. This is essentially why this formalism accommodates 
all nice features of perturbative field theory, 
just curing its single but giant defect, namely divergence.

The most aesthetic and compact 
formulation of perturbative QFT is the parametric representation,
and it is also the one in which space time is no longer at the center of the stage.
The associated idea of dimensional interpolation is a beautiful feature of perturbative 
QFT which was essential in two key milestones in the development of QFT:
the proof of renormalizablity of non Abelian gauge theories by 'tHooft and Veltmann 
\cite{HV} and the Wilson-Fisher $\epsilon$ expansion \cite{Wil}. 
These are certainly milestones to which one would like to give 
constructive meaning. Parametric representation relies on
various types of tree matrix \cite{A1} or tree Pfaffian theorems \cite{GR,RT}. This again 
points towards trees as the fundamental structure in QFT.

The good news is that our formalism is especially compatible with that parametric representation,
to the point that it could be described as a kind of ''constructive parametric'' formalism.
Indeed the canonical forest or tree formula can be adapted so that
its corresponding interpolating parameters just coincide with a subset
of Feynman-Schwinger parameters, those for the tree considered!
In this way tree matrix elements of the scalar products corresponding to QFT models 
become just finite sums of finite dimensional Feynman integrals in parametric space,
with just some funny new condition on the range of integration
of the ``loop parameters''. These conditions are now a really small cost 
to go from perturbative to constructive QFT!

Remark that Fermionic field theory has undergone quietly this tree revolution almost two decades ago.
After a long period of maturation \cite{Les,FMRT1,AR2}, it lead to 
full constructive results such as the rigorous definition of 
differential renormalization group equations for Fermions \cite{DR1} 
and to a flurry of theorems on condensed matter \cite{DR2,FKT,Hub, BGM}. 
However the full power of the idea of basing QFT on trees was still not 
recognized at that time because Bosonic theories could not be brought into that form.

The stimulation for finding this better formalism came from an unexpected 
source, namely the discovery of a simple quantum field theory on the four 
dimensional Moyal-Weyl space, the Grosse-Wulkenhaar model \cite{GW}. That model is 
renormalizable \cite{GW,GW2,RVW,GMRV} and 
asymptotically safe \cite{GrWubeta,DRbeta,DGMR}. It is therefore an extremely tempting target, 
as first potential example of a simple and mathematically well-defined non trivial four dimensional QFT\footnote{Yang-Mills theory is neither simple, nor yet fully well defined in four dimensions.}.

However the (rather ugly, one must admit) technique of multiscale cluster and Mayer 
expansions \cite{GJ, Riv1,Br,AR3}, the only available constructive tool for Bosonic quantum 
field theories, could not be applied to the Grosse-Wulkenhaar model, essentially 
because the interaction of that model is non-local in $x$-space. In matrix
base the problem is simply shifted:  cluster expansions apply to vector models but not to
non-trivial matrix models (they do not provide the right bounds 
when the size of the matrix increases). Hence something better
had to be found.

A first progress occurred one year ago when one of us
found that combining the canonical forest formula with the intermediate field method
lead to a convergent resummation for matrix models
uniformly in the size of the matrix \cite{R1}. The resulting
loop vertex expansion was devised to treat 
a renormalization group slice for the Grosse-Wulkenhaar model, and
it was quickly realized that this method applied to ordinary
quantum field theory on commutative space as well \cite{MR1}.
We recommend the reading of \cite{R1,MR1} before going further down this paper.
However three main drawbacks remained:

\begin{itemize}

\item
The intermediate field method does not generalize easily to other stable
interactions than $\phi^4$, say $\phi^6$ etc... It is probably possible to treat these other cases
with more and more intermediate fields  but the technique becomes cumbersome.

\item
A second problem (in fact deeply related to the first)
is that the loop vertex expansion of \cite{R1,MR1} is not easily ordered into 
a multiscale expansion suited for the renormalization group. 
This is because for instance a loop made of propagators
of two different scales does not factor as two loops, one in each scale,
but rather as a sequence of open single-scale resolvents. Although again
some ways to circumvent this difficulty do exist, they are not elegant.
The conclusion is that some kind of resolvent, rather than loop, 
should be the right canonical object.

\item
Finally in the loop vertex expansion, functional integration is still present,
although in the reduced, more "model-independent" form of white noise
for the intermediate $\sigma$-field.
Therefore the formalism does not seem to 
lead to new insights on QFT in non-integer dimension.
Formulating quantum field theory in noninteger dimension 
is a key benchmark to supersede perturbative field theory while retaining its advantages.
\end{itemize}

The formalism developed in this paper solves the first two problems at once,
and leads to a new way of attacking the third.
The vision of QFT that emerges is that of a generating functional for the
species of weighted trees that automatically computes 
connected functions\footnote{Species are 
roughly speaking structures on finite sets of points, together with generating functionals 
which allow to extend usual operations on functions, 
and therefore to formulate rigorously statements such as the
logarithm of forests are trees, the derivatives of cycles are chains etc...} \cite{BLL}.

Constructive bounds now reduce essentially to the positivity of the 
universal Hamiltonian operator. The vacuum is the trivial tree and  
the correlation functions are given by  "vacuum expectation values" of the resolvent
of that combinatoric Hamiltonian operator. The resulting formulation of the theory is
given by a convergent rather than divergent expansion. In short and at the
most naive level, this is because there are much less trees than graphs, but they still
capture the vital information about connectedness.

We have worked out our method in the test case of a $\lambda \phi^4$ 
interaction with a real coupling constant $\lambda$
and in a real dimension $D$. 
Let us remark that in all known cases where QFT has been built, 
perturbative theory was Borel summable. We conjecture that this is indeed
the case and that our non-perturbative definition of QFT 
is indeed the Borel sum of ordinary perturbation theory.
We formulate a conjecture on positivity of certain
matrices which would allow to also extend 
the theory to real non-integer dimensions, and probably also to complex dimensions 
with positive real parts. This would be an important step towards
making precise the mathematical status of the Wilson-Fisher expansion.

The essential point is that in this reformulation of QFT space-time no longer 
lies at the center of the stage. Topological notions such 
as trees now play that central role.  Therefore
we hope this point of view might ultimately help 
to answer deep questions, such as:  Is quantum gravity a quantum field theory 
in some "extended sense"? Is it renormalizable?

This paper should be really thought as an introduction to a new line of ideas. 
The mathematical core of the paper is in section 4 where we give in detail
positivity theorems. More precise mathematical details and the exploration 
of the many conjectures that this work suggests, in particular those of the last sections
are devoted to future publications.

\medskip\noindent{\bf Acknowledgments}

We thank A. Abdesselam and P. Leroux for the organization 
of a very stimulating workshop on combinatorics 
and physics, which was one of the many sources of inspiration
for this paper.

\section{General Formalism}
\setcounter{equation}{0}

\subsection{The Universal Vector Space}

The basic quantities of field theories, the $N$-point correlation
functions, are described by sums of connected graphs and are functions
of a certain set of external invariants. In the parametric representation
we know that the dependence in terms of the Euclidean invariants 
is associated to a sum over two-trees of the graph, which are 
similar to spanning trees but with one special deleted or cut line.
The invariant associated to such a two tree is the square of the
sum over all incoming momenta on any of the two
pieces defined by the cut. It can be computed on any of them because
by momentum conservation the corresponding invariants are equal.
Note also that cuts which don't have any incoming line on one piece
don't contribute as their associated invariant is 0.

Motivated by this observation we introduce the family of labeled\footnote{
We use labeled trees because they are the most standard ones. 
Labeled means that vertices are labeled; the total
number of labeled (unmarked)  trees with $n$ vertices and $n-1$
lines is $n^{n-2}$ (Cayley's theorem).}
marked trees with one external point or source and one mark, both of which are leaves.
The order $n$ of such a tree is defined as the number of vertices (excluding the mark and the external point).
To the source vertex will be associated an external variable, eg a spatial position $x$. The line with the mark at one end 
is special and should be thought as a half-line, waiting to be glued to another one of the same type.

\begin{figure}[htbp]
\centering
\includegraphics[scale=.5,angle=270]{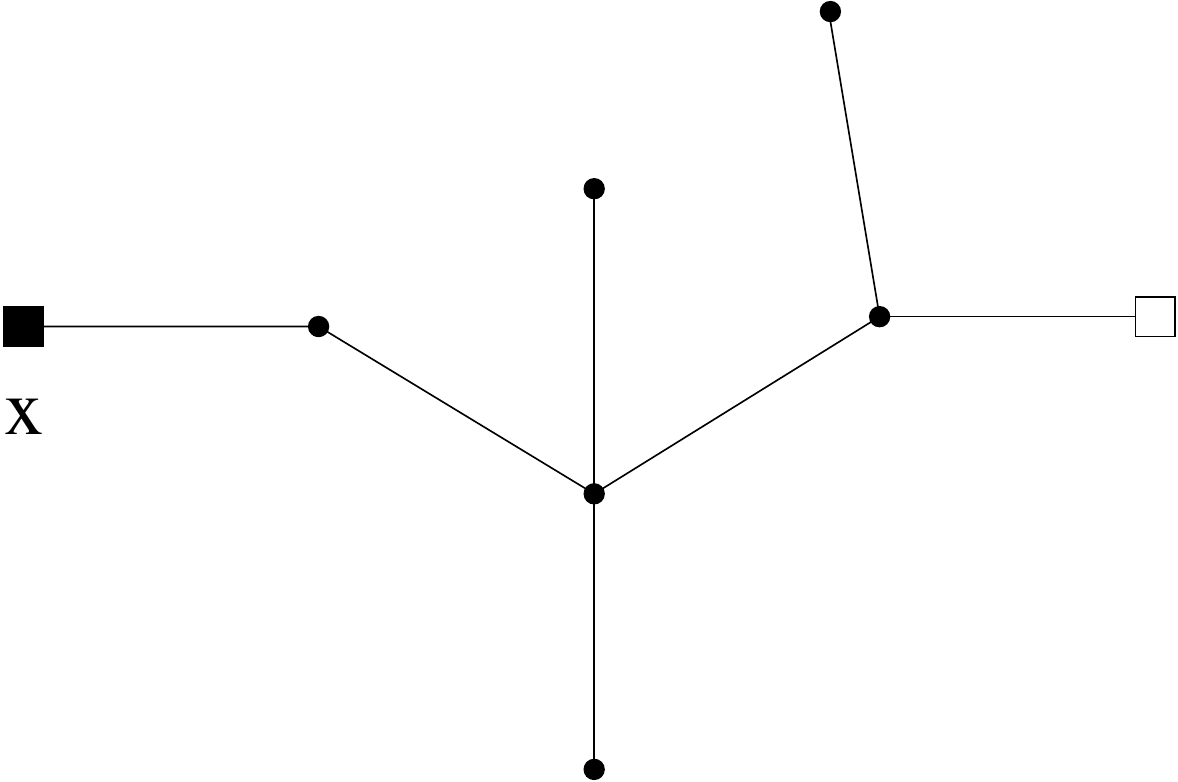}
\caption{A marked tree of order n=6. The mark is the white box; the black box is the source at position $x$.}
\label{markedtree}
\end{figure}

The universal vector space for QFT is an (infinite dimensional) 
vector space $\cE$ which is the algebraic vector space spanned by the countable
basis $e_T$ for each such marked tree $T$. It foliates as 
\be  \cE = \oplus_{N \ge 1}  \cE_N
\ee
where $\cE_N$ is spanned by marked trees with $N$ sources. It contains a natural
exhausting sequence of finite dimensional spaces if we fix eg the total number of lines
of the tree. Finally we recall that each element of $\cE$ is a linear combination
\be  \sum_T  \lambda_T  e_{T}
\ee
where the sum over $T$ is \emph{finite}.

This universal space will be decorated by a (model dependent)
scalar product and will then give rise under completion 
to various (model-dependent) Hilbert spaces $\bar \cE$.

\subsection{The Universal Hamiltonian}

Interesting operators on $\cE$ may be obtained by operations such as gluing or contracting lines of trees. 
We focus on one particular operator 
which plays the key role in what follows and which we call the "universal Hamiltonian".
It is in fact only really defined in $\bar \cE$, but it is an inductive limit of a family of 
operators $H_n$ defined in $\cE$ as follows.

We need first to introduce a new category of trees called elementary 2-marked trees. They have no sources, two
special marked leaves, and the property that the (unique) path from one gluing point to the other one
contains \emph{exactly one vertex}.

\begin{figure}[htbp]
\centering
\includegraphics[scale=.5,angle=270]{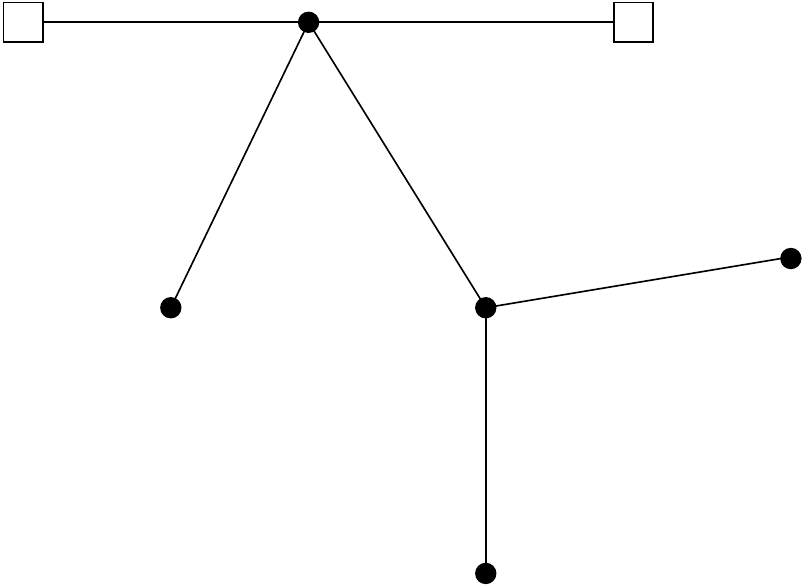}
\caption{An elementary 2-marked tree of order n=5.}
\label{eltree}
\end{figure}

There is a natural gluing operation of a marked tree (with $N$ sources) $T$ and an elementary two-marked tree
$S$. It creates a larger marked tree (with $N$ sources) $S\star T$. It glues the marked point of $T$
to one of the marked point of $S$ fusing their (half)-lines into a single line.

We now define the $n$-th order universal or "abstract"  Hamiltonian $H_n$ by its action on basis vectors
$H_n e_T = -\sum_{S,  n(S) \le n}  e_{S\star T}$
where $n(S)$ is the order of $S$, in which marks do not count.

\medskip

\noindent {\bf Remarks}

\begin{itemize}

\item The definition of the gluing does not depend on which mark 
we chose in $S$ for the gluing.

\item The sum over $S$ being finite, the operators $H_n$ are well-defined
on $\cE$. 

\item  $lim_n H_n$ does not exist on $\bar \cE$, but $   lim_n  (1+H_n)^{-1}$ will exist.

\end{itemize}

\begin{figure}[htbp]
\centering
\includegraphics[scale=.5,angle=270]{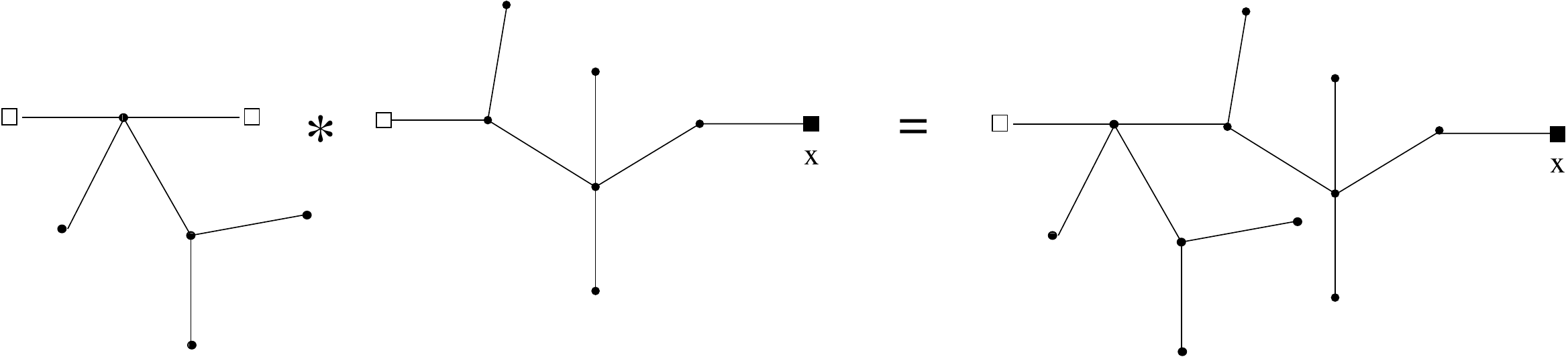}
\caption{The gluing operation.}
\label{glutree}
\end{figure}

This completes the list of universal model-independent structure.  Of course when really changing
QFT, eg to NCQFT the categories have to change, eg fermions imply oriented trees,
NCQFT imply ribbon trees\footnote{For ribbon graphs, the
$\star$ operation is not symmetric but 
$H_n$ remains symmetric.} and so on.

\subsection{The Forest Formula}

Consider $n$ points; the set of pairs $P_n$ of such points which has
$n(n-1)/2$ elements $\ell = (i,j)$ for $1\le i < j \le n$, and a smooth function $f$
of $n(n-1)/2$ variables $x_\ell$, $\ell \in \cP_n$. Noting $\partial_\ell$ 
for $\frac{\partial}{\partial x_\ell}$, the standard canonical forest formula is \cite{AR1}

\be\label{treeformul1}
f(1,\dots ,1)
= \sum_{\cF}  \big[ \prod_{\ell\in \cF}   
\int_0^1 dw_\ell   \big]� \big( [ \prod_{\ell\in \cF} \partial_\ell ] f 
\big)  [ x^\cF_\ell (\{ w_{\ell'}\} ) ]
\ee
where 
\begin{itemize}

\item the sum over $\cF$ is over forests over the $n$ vertices, including the empty one 

\item $x^\cF_\ell (\{ w_{\ell'}\} )$ is the infimum of the $w_{\ell'}$ for $\ell'$
in the unique path from $i$ to $j$ in $\cF$, where $\ell = (i,j)$. If there is no such path,
$x^\cF_\ell (\{ w_{\ell'}\} ) = 0$ by definition.

\item The symmetric $n$ by $n$ matrix $X^\cF (\{w\})$ defined
by $X^\cF_{ii} = 1$ and $X^\cF_{ij} =x^\cF_{ij} (\{ w_{\ell'}\} ) $ 
for $1\le i < j \le n$ is positive.

\end{itemize}

A particular variant of this formula (\ref{treeformul1})  is in fact
better suited to direct application to the parametric representation 
of Feynman amplitudes. It consists in changing variables $x \to 1-x$
and rescaling to $[0,1] \to [0,\infty]$ of the range of the variables.
One gets that if $f$ is smooth with well defined 
limits for any combination of $x_\ell$ tending to $\infty$,

\be\label{treeformul2}
f(0, \dots, 0) = \sum_{\cF}   \big[ \prod_{\ell\in \cF}   
\int_0^\infty  ds_\ell   \big]� \big( [ \prod_{\ell\in \cF} - \partial_\ell ] f 
\big)  [ x^\cF_\ell (\{ s_{\ell'}\} ) ]
\ee
where 
\begin{itemize}

\item the sum over $\cF$ is like above,

\item $x^\cF_\ell (\{ s_{\ell'}\} )$ is the \emph{supremum} of the $s_{\ell'}$ for $\ell'$
in the unique path from $i$ to $j$ in $\cF$, where $\ell = (i,j)$. If there is no such path,
$x^\cF_\ell (\{ s_{\ell'}\} ) = \infty$ by definition. This is because the change of variables
exchanged $\inf$ and $\sup$.
\end{itemize}

To distinguish these two formulas we call $w$ the parameters of the first one (like "weakening")
since the  formula involves infima, and $s$ the parameters of the second one (like "strengthening" or "supremum")
since the formula involves suprema.

\section{QFT Models as scalar products}
\setcounter{equation}{0}

A QFT model is defined perturbatively by gluing propagators and vertices and computing 
corresponding amplitudes according to certain Feynman rules.

Model dependent features imply the types of lines (propagators for various
particles, bosons or fermions), the type of vertices,
the space time dimension and its metric (Euclidean/Minkowsky, curved...)

We now reorganize the Feynman rules
by breaking Feynman graphs into pieces and putting them into boxes
labeled by trees according to the canonical formula.
The content of each tree box is then considered the matrix element
of a certain scalar product.

Equivalently we can obtain our formalism by applying the canonical forest  formula
to the $n$-th order of perturbation on a functional integral. Of course we arrive at the same point.
However one can use either formula (\ref{treeformul1}), introducing  new weakening parameters $w$,
or formula  (\ref{treeformul2}) in which the $s$ parameters are directly the Feynman parameters
of the parametric representation of the tree lines. This second point of view is much better 
suited to multiscale analysis, but for pedagogical reasons
we give both formulas. We  illustrate our formalism with the example of the $\lambda \phi^4$ theory
in real  space time dimension $D$. Other stable polynomial interactions
could be treated in the same way. 

\subsection{Propagator}
For Bosonic scalar field theory in integer dimension the usual massive propagator is,
up to inessential constants, which we forget from now on
\bea
C(k) &=& \frac{1}{k^2 + m^2}  = \int_0^{\infty} e^{- \alpha (k^2 + m^2)} d\alpha \\
C(x,y) & =& \int_0^{\infty}  \frac{e^{-\alpha m^2 - \vert x-x' \vert ^2 / 4\alpha }}{\alpha^{D/2}} d\alpha 
\eea

We also note $D$ the propagator at fixed value of the Feynman parameter
\be D(s; x,x') = \frac{e^{-sm^2 - \vert x-x' \vert ^2 / 4s}}{s^{D/2}} 
\ee

\subsection{Tree Amplitudes}

We consider now ordinary trees with $N$ external points, $N\ge 1$. To any such tree
we shall associate an amplitude by applying formulas (\ref{treeformul1}) or (\ref{treeformul2})
to the functional integral defining the theory, which is only at this stage a heuristic tool.
The goal is to obtain a forest formula for unnormalized functions, from which
a tree formula for normalized connected functions follows. 

We show now two ways to apply these formulas. 
As a pedagogical exercise in subsection \ref{subsec1}� we use (\ref{treeformul1}) 
to decouple vertices in the most naive way. This does not 
optimize multiscale analysis.

Then we show in subsection \ref{subsec2} how to apply the second formula  (\ref{treeformul2})
directly on the Feynman parameters in the parametric representation,
This has two main advantages. First  Feynman parameters precisely provide scale analysis,
so that  we obtain a much better formalism
for future applications in which renormalization will enter the picture.
Second, in the parametric representation space time is a parameter
and we get in this way a program to define QFT 
constructively at non integer dimension, which we sketch in section \ref{nonint}.

\subsubsection{Scale independent amplitudes}\label{subsec1}

The first (naive interpolation) computes an amplitude (in $x$ space representation)

\bea  A(T, z_1, \dots ,z_N) &=&\frac{(-\lambda)^n}{(4!)^n n!} \int  \prod_{v \in T}  dx^D_v   \prod_{\ell \in T}  \int_0^1 d w_{\ell}
\prod_{\ell \in T}  C (x_{\ell}, x'_\ell)
\\  \nonumber
&&\int d\mu_{C^{T}(w)}
\biggl(  \prod_{\ell \in T}  \frac{\delta}{\delta \phi(x_\ell)} \frac{\delta}{\delta \phi(x'_\ell)}   
 \prod_{i=1}^N \phi(z_i)   \prod_v \phi^4(x_v)  \biggr)
\eea
where $C^T(x,x', w) =  C(x,x') \inf_{l \in P^T(x,x')} w_{l}$, and $P^T(x,x')$ is the path 
in $T$ from $x$ to $x'$. The functional derivations $\prod_{\ell \in T}  \frac{\delta}{\delta \phi(x_\ell)} \frac{\delta}{\delta \phi(x'_\ell)}  $ of course are constrained to apply in such a way as to create 
the lines of the tree $T$.
This formula comes by applying the interpolation formula (\ref{treeformul1}) where
a distinct field copy is associated to each vertex with a degenerate copy-blind covariance. 
This copy-blind covariance is weakened on off diagonal terms by the $w$ factors.
{This is exactly the same method that was used in \cite{R1} for the $\sigma$ field.}

\medskip\noindent{\bf Remark 1} 

We recall that such amplitudes are distributions in the external arguments 
$z_1, .... z_N$ which may be singular at coinciding points.
We can smear them against test functions $f_1, \dots f_N$.
It is well known that bare Feynman amplitudes may diverge for $D$ large enough.
This is tackled through renormalization theory, which has of course to
enter the picture when necessary.  Here and in what follows 
the reader may assume $\Re D < 2$ to separate the issues.

\begin{figure}[htbp]
\centering
\includegraphics[scale=.5,angle=270]{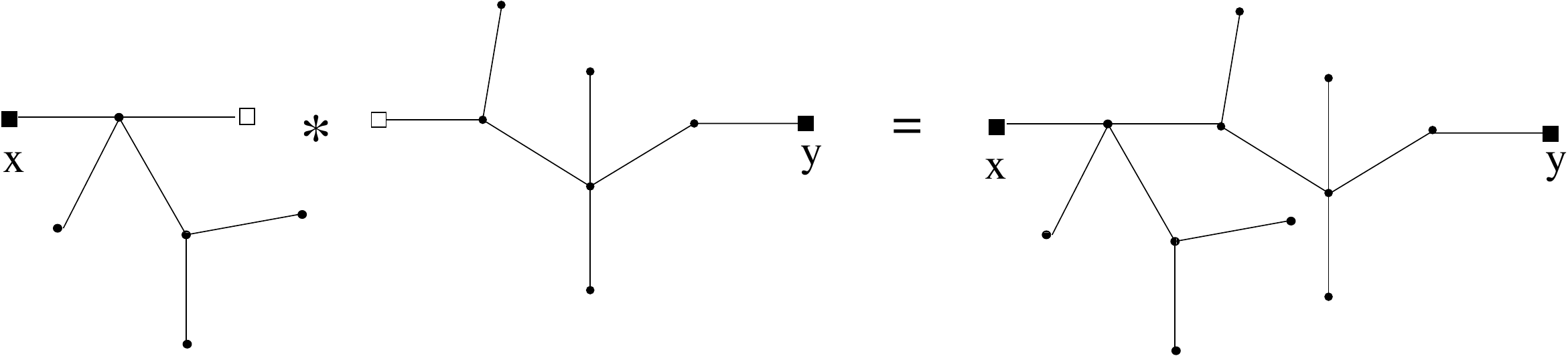}
\caption{The gluing of two marked trees into an ordinary tree}
\label{treeglued}
\end{figure}

\medskip\noindent{\bf Remark 2} 
Working with the $\phi^4$ means that we cannot produce any tree with degree more than
4 at any vertex. Therefore we could restrict the space ${\cal E}$ with that condition. 
We prefer to consider that all amplitudes for trees which violates that condition are zero.
The amplitude for a tree such as the one of Figure \ref{treeglued} should be thought as obtained
by first completing all vertices of degree less than 4 to degree 4 by adding the necessary fields, i.e. half lines, then summing over all their contraction schemes with the correct weakening parameters.

\subsubsection{Parametric amplitudes}\label{subsec2}

This method was introduced for Fermions in  \cite{DR1}.

\bea  A(T, z_1, \dots ,z_N) &=&\frac{(-\lambda)^n}{(4!)^n n!} \int  \prod_{v \in T}  dx^D_v   \prod_{\ell \in T}  \int_0^{\infty} 
d s_{\ell}
\prod_{\ell \in T} [ D (s_\ell ; x_{\ell}, x'_\ell) ] \nonumber\\
&&\int d\mu_{C^{T}(s)}
\biggl(  \prod_{\ell \in T}  \frac{\delta}{\delta \phi(x_\ell)} \frac{\delta}{\delta \phi(x'_\ell)}   
 \prod_{i=1}^N \phi(z_i)   \prod_v \phi^4(x_v)  \biggr) \nonumber\\
&&\eea
where 
\be
C^{T}(s; x,x' ) = 
\int_{\sup_{l \in  P^T(x,x') } s_l}^{\infty}   \frac{e^{-\alpha m^2 - \vert x-x' \vert ^2 / 4\alpha }}{\alpha^{D/2}} d\alpha 
\ee
is the propagator with a restricted integration range in parametric space.
This is obtained by applying formula (\ref{treeformul2}) directly to the Feynman parameters.

\subsection{The scalar product}

There is a natural  gluing operation $\star$
on marked trees with sources. To  two such marked trees $T$ and $T'$
with  $p$ and $q$ sources it associates an ordinary tree $T \star T'$ with
$p+q$ sources, by gluing the two marked (half lines) into an ordinary line (always called
$\ell_0$ in what follows).

We now consider the  infinite matrix
$<e_T, e_{T'}> = A(T \star T')$. This matrix is obviously symmetric
because $T\star T' = T' \star T$. 

\begin{theorem}
This matrix is  positive, hence defines a scalar product on $\cE$
\end{theorem}

The theorem means that
$< \sum_T \lambda_T   e_T  ,  \sum_T \lambda_T   e_T > \ge 0$, $\forall \lambda_T$, the sum
being over finitely many marked trees.

The operator $H_n$ is symmetric with respect to that scalar product, because

\bea  <e_T , H_n e_{T'}>  &=&  <e_T, -\sum_{S,  n(S) \le n}  e_{S\star T'}> 
\nonumber\\&=& -\sum_{S,  n(S) \le n}  A(T \star ( S \star T') )
= <H_n e_T , e_{T'}>
\label{gluesym}
\eea
because
\begin{equation} T \star ( S \star T') =  (S \star T) \star T' 
\label{gluesym2}
\end{equation}
Remark that for ribbon graphs, equation (\ref{gluesym2}) is not true
but (\ref{gluesym}) still holds because of the summation.

\begin{theorem}
The operator $H_n$ is positive, i.e. 
$< \sum_T \lambda_T   e_T  , H_n ( \sum_T \lambda_T   e_T )> \ge 0$, $\forall \lambda_T$, $\forall n$, the sum
being over finitely many marked trees.
\end{theorem}

The proofs of these two main theorems is  given in the following section.

This positivity is a kind of abstract tree version of the well known OS Euclidean positivity axiom. 
Developing this analogy should lead to an axiomatic formulation of Tree QFT that we leave 
for the future.

\section{The positivity theorems}
\setcounter{equation}{0}

\subsection{Positivity of the scale independent formula} 

Let us call $z_T$ the collection of fixed external positions
$z_1, ..., z_N$ of a marked tree $T$ with $N$ sources,
and $f_T (z_1,...,z_N)$ a test function for these sources.
We want to prove that 

\be I =  \sum_{T, T'} \lambda_T   \lambda_{T'} \int dz_T dz_{T'} 
A(T\star T', z_T, z_{T'})  f_T (z_T)  f_{T'} (z_{T'}) \ge 0
\ee
hence that 
\bea I &=& \sum_{T, T'} \lambda_T   \lambda_{T'} \int dz_T dz_{T'}  f_T (z_T)  f_{T'} (z_{T'})
\frac{(-\lambda)^{n(T\star T' ) }}{(4!)^{n(T \star T')} n(T \star T')!}
\nonumber
\\ && \int  \prod_{v \in T\star T'}  dx^D_v   \prod_{\ell \in T\star T'} 
 \int_0^1 d w_{\ell}
\prod_{\ell \in T\star T'}  C (x_{\ell}, x'_\ell)  \int d\mu_{C^{T\star T'}(w)}   \nonumber
\\
&&
\biggl(  \prod_{\ell \in T\star T'}  \frac{\delta}{\delta \phi(x_\ell)} \frac{\delta}{\delta \phi(x'_\ell)}   
\prod_{i=1}^{N(T \star T')} \phi(z_i)   \prod_{v \in T \star T' } \phi^4(x_v)  \biggr)  \ge 0 \label{bigsumT}
\eea

We reorganize this sum by fixing the number $p \ge 0$ of lines which cross 
from $T$ to $T'$ in the functional integration above, and we prove that for any fixed
$p$ the sum above is positive. 

The gluing line $\ell_0$ in $T\star T'$ has by convention index 0 and associated weakening parameter $w_0$.  It has
propagator $C(x_0, x'_0)$.
The other crossing lines have indices $i = 1, \cdots p$.
We use the identity $\inf (w)  = \int_0^{\inf w} du $
to express as integrals over new parameters $u_i$ the weakening 
parameters of the $p$ crossing lines.
We use the multinomial identity plus relabeling to attribute vertices either to the right or 
to the left of the star operation.
This replaces the $1/n(T \star T')!$ symmetry factor by "factorized" factors $1/n(T) ! n(T')!$. Finally for $i=0, \cdots p$ 
we cut the crossing lines $C(x_i, x'_i)$ in the middle with respect to new variables
$y_i$, as $C(x_i, x'_i) = \int d^Dy_i C^{1/2}(x_i, y_i) C^{1/2}(y_i, x'_i)  $.
This rewrites $I$ in formula (\ref{bigsumT})
as
\bea I =\sum_{p \ge 0}  \frac{1}{p!} \int_0^1 dw_0   \prod_{i=1}^p  \int_0^{w_0}  du_i    \prod_{i=0}^p \int  d^Dy_i 
\biggl(  K_p(\{y\} , \{ u\} )    \biggr)^2
\label{kernel}
\eea
\bea
K_p(\{y\}  \{ u\}) =
\sum_{T}  \frac{  \lambda_T  (-\lambda)^{n(T)}  }{    (4!)^{n(T)} n(T)! }     \int dz_T  f_T (z_T)
  \int  \prod_{v \in T}  dx^D_v  \prod_{\ell \in T} 
\int_0^1 d w_{\ell}
\\
 C^{1/2} (y_0, x_0) \prod_{\ell \in T}  C (x_{\ell}, x'_\ell)
\int d\mu_{C^{T}(w)}  \biggl(  \prod_{i=1}^p  \int d^D x_i C^{1/2} (y_i, x_i) \frac{\delta}{\delta \phi(x_i)}    \biggr) \nonumber
\\ 
 \biggl(  \frac{\delta}{\delta \phi(x_0)}   \prod_{\ell \in T}  \frac{\delta}{\delta \phi(x_\ell)} \frac{\delta}{\delta \phi(x'_\ell)}   \biggr)
\prod_{j=1}^{N(T)} \phi(z_j)  \prod_v \phi^4(x_v)   \prod_{i=1}^p \chi ( w^T_i  \ge u_i )   \nonumber
\eea
where $w^T_i $ is the infimum over the parameters in the unique path in $T$ 
going from $x_i$ to $x_0$. Note that the non trivial function  
$ \prod_{i=1}^p \chi (\inf w^T_i  \ge u_i ) $  can be computed only after
the action of the functional derivatives, as it depends on this action. The important point
is that the condition that the crossing 
lines are multiplied by the infimum of the $w$ over the path in $T \star T'$
can be factorized in these non trivial functions, thanks to the $u$ parameters. This
trick is a multiparameter generalization of the identity
\bea
\int_0^1 \int_0^1   dx dy  \inf (x,y) f(x) f(y) = \int_0^1  ds  \int_s^1 \int_s^1   dx dy f(x) f(y)  \ge 0
\eea

From (\ref{kernel}) follows immediately that $I \ge 0$.

Finally the positivity for $H$ can be proved exactly in a similar manner,
but we have to split the central vertex in two halfs. In short the role of the 
$C_0$ line is replaced by a propagator which is in fact a delta function 
and there is no $w_0$ parameter. Apart from these details the factorization is identical.
Remark that this split of $H$ as a square can be performed for any even polynomial,
not only $\phi^4$. This is definitely an advantage of this method over the intermediate field 
method and loop vertex expansion of \cite{R1}-\cite{MR1}.

\begin{figure}[htbp]
\centering
\includegraphics[scale=.5,angle=270]{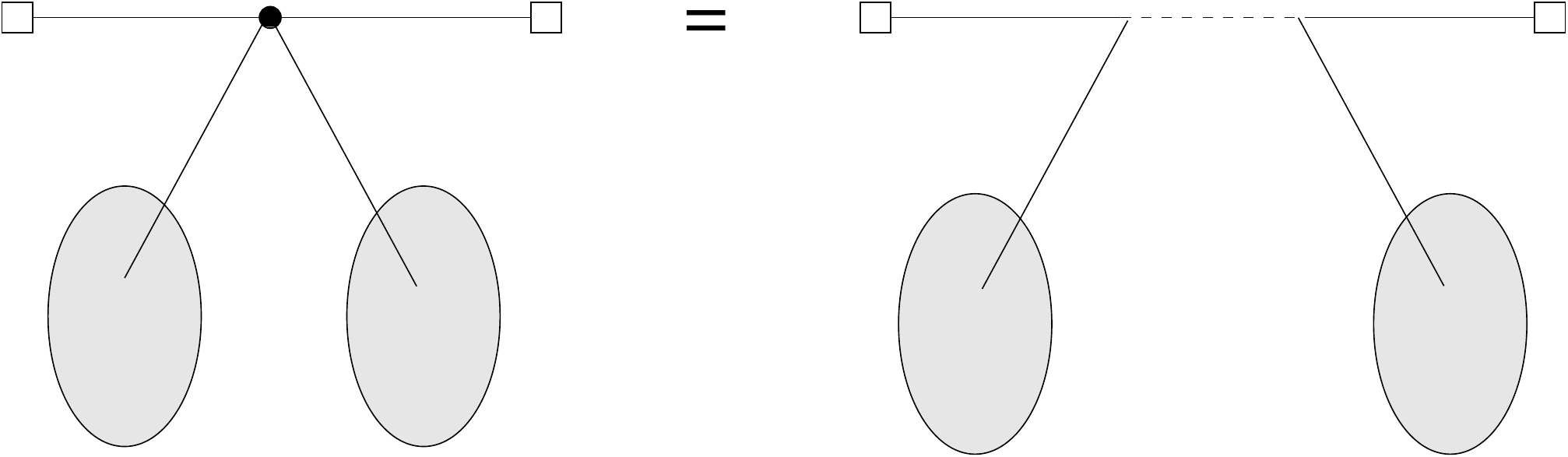}
\caption{Positivity of the $H$ operator. The dotted line represents a delta function.}
\label{deltasym}
\end{figure}

\subsection{Positivity of the parametric formula}

We use now formula (\ref{treeformul2}). The outcome is only slightly different, and reads for 
an integer dimension $D$:

\bea I =\sum_{p \ge 0}  \frac{1}{p!} \int_0^\infty ds_0    
\prod_{i=1}^p   \int_{s_0}^\infty dt_i  \prod_{i=0}^p   \int  d^Dy_i 
\biggl(  K_p(s_0, \{y\} , \{ t\} )    \biggr)^2
\label{kernel1}
\eea
\bea \label{parainteger}
K_p(s_0, \{y\} ,  \{ t\}) &=& 
\sum_{T}  \frac{  \lambda_T  (-\lambda)^{n(T)}  }{    (4!)^{n(T)} n(T)! }     \int dz_T  f_T (z_T)
\int  \prod_{v \in T}  dx^D_v \\
&& \prod_{\ell \in T} 
\int_0^\infty d s_{\ell} D^{1/2} (s_0 ; y_{0}, x_0)
\prod_{\ell \in T}  [ D (s_\ell ; x_{\ell}, x'_\ell) ] \int d\mu_{C^{T}(s)}  \nonumber
\\
&&\biggl(  \prod_{i=1}^p  \int d^D x_i [D(t_i; y_i, x_i)]^{1/2} \frac{\delta}{\delta \phi(x_i)}    \biggr)  \nonumber
\\
&&
\biggl( \frac{\delta}{\delta \phi(x_0)}   \prod_{\ell \in T}   \frac{\delta}{\delta \phi(x_\ell)} \frac{\delta}{\delta \phi(x'_\ell)}   \biggr) \nonumber
\\ 
&&
\prod_{j=1}^{N(T)} \phi(z_j)  \prod_v \phi^4(x_v)   \prod_{i=1}^p \chi ( s^T_i  \le t_i )   \nonumber
\eea
where $s^T_i $ is the supremum over the parameters in the unique path in $T$ 
going from $x_i$ to $x_0$. 

The proof that $H$ is positive is similar.

\section{The Hilbert space and the two point function}
\setcounter{equation}{0}

The Hilbert space is defined as the completion $\bar \cE$ of the canonical space for the 
chosen scalar product. It is therefore model dependent. We already 
know from Haag's theorem that this should be the case. Our scalar product typically (for instance for the $\phi^4$ theory)
has matrix elements of order $n!$ so that the Hilbert space is 
roughly made of infinite sums of trees with coefficients decaying as $1/n!$.

The full abstract Hamiltonian operator $H$  is  the inductive limit $\lim_{n \to \infty} H_n$.
It is not well defined on $\cE$ nor on the whole of $\bar \cE$, because it is an unbounded operator.
However the interacting propagator
$\frac{1}{1+H}$ can be defined as $\lim_{n \to \infty}  \frac{1}{1+H_n}$.
and should be a bounded operator on $\bar \cE$.

An analogy that may help to grasp the situation is that of the ordinary Laplacian;
although $\Delta$ is not a bounded operator on $L^2$, we can perfectly
define $(1 - Delta)^{-1}$ on that space.

\begin{definition}
The interacting propagator or two point function is then defined non-perturbatively by
\begin{equation}  \label{maindef2point}
S_2 (x,y)= <e_0 (x) , \frac{1}{1+H} e_0 (y)>
\end{equation}
\end{definition}
where $e_0 (x) $ is the trivial tree whose black box position is at $x$.

We do not give details here about existence of the limit $\frac{1}{1+H}= \lim_{n \to \infty}  \frac{1}{1+H_n}$ 
but it should follow easily from the positivity of all $H_n$'s, and lead to the norm
bound $\Vert  \frac{1}{1+H} \Vert \le 1$. To establish the decay properties of $S_2 (x,y)$
as $\Vert x - y \Vert \to \infty$ however is expected to require expansion steps followed by 
inequalities similar to those of \cite{MR1}.

\section{$N$ point functions}
\setcounter{equation}{0}

It is also possible to define $N$ point functions, but it requires to enlarge slightly
the formalism. Formally the $N$ point functions can be obtained
by gluing two marked trees with $p$ and $q$ sources with $p+q =N$>
However the crucial non-perturbative ingredient is hidden in  the positivity of $H$
 and the resolvent $(!+H)^{-1}$ in (\ref{maindef2point}). To make use also of
this resolvent, we can define for any $N$ point function its skeleton, which is made of
of at most $N-2$ particular "crossroad" vertices $V_c$ and of thick lines.

\begin{figure}[htbp]
\centering
\includegraphics[scale=.45,angle=270]{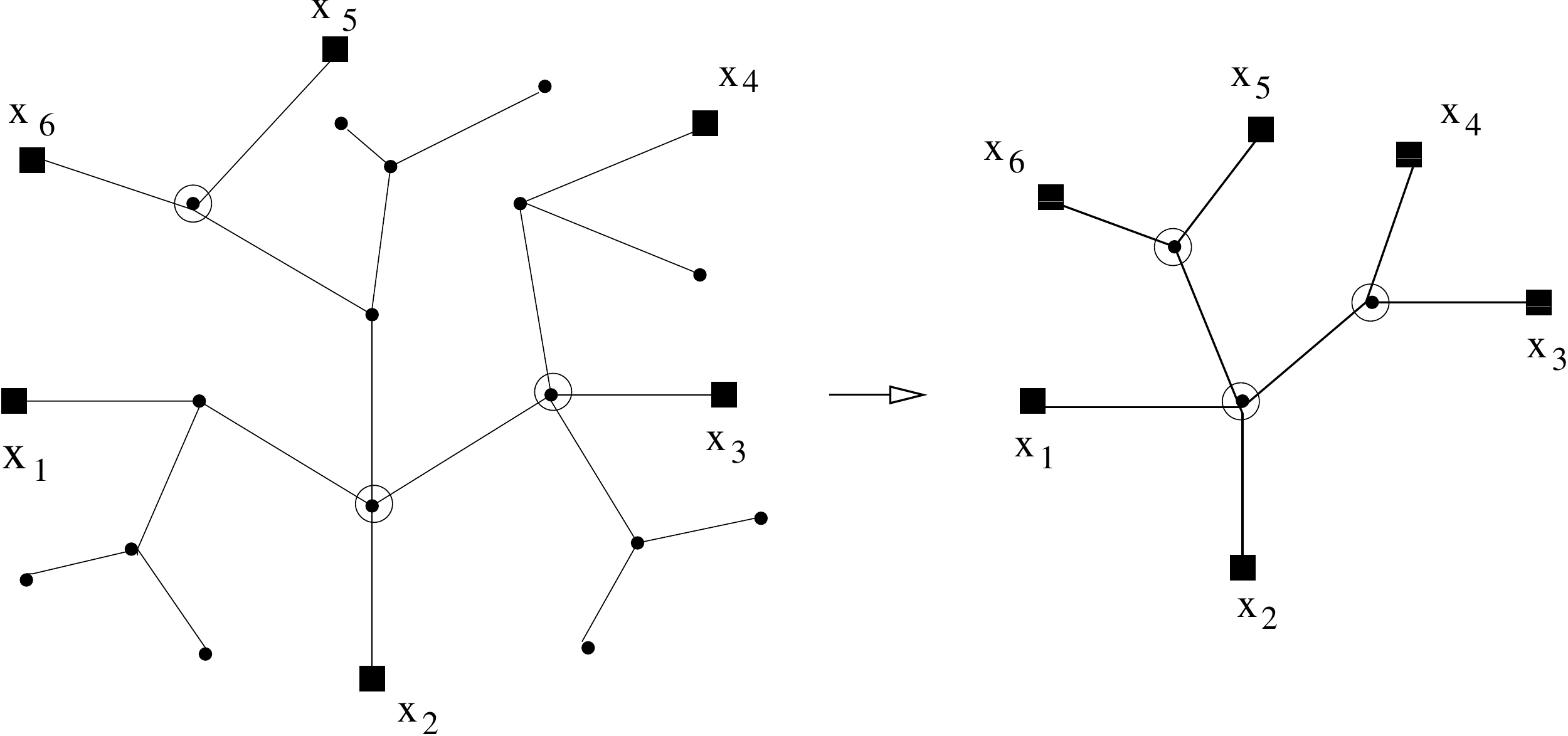}
\caption{The skeleton of a tree with $N=6$ sources and three "crossroads".}
\label{skel}
\end{figure}

The thick lines correspond
to resolvents $(1+H)^{-1}$. A crossroad vertex of degree $d$ can be though as having 
$d$ gluing marks or white boxes, and an associated coupling constant $-\lambda$; in a $\phi^4$ 
theory we must again have $d \le 4$. In the universal space $cE$ such a vertex is not an operator
but an operad, that it is tensor of degree $d$.
The  non perturbative definition of $N$
point Schwinger functions is then

\begin{definition}
The interacting $N$ point functions are defined non-perturbatively by
\begin{equation}  \label{maindefNpoint}
S_N (x_1, .... x_N) = \sum_{N-skeleton  \cS}   \prod_{v \in \cS} V_v   \prod_{\ell \in \cS} \bigl( \frac{1}{1+H} \bigr)_\ell 
\prod_{j=1}^N   e_0 (x_j)
\end{equation}
\end{definition}
where the gluing are made according to the skeleton $\cS$ as in Figure \ref{skel}, with hopefully transparent notations.
The important point is that the sum over $\cS$ is {\it finite}, hence this is a convergent definition, as all divergences
of ordinary perturbation theory have been hidden in the  $( \frac{1}{1+H})_\ell$ resolvents.

A few additional remarks are in order.

Like the loop-vertex expansion this formalism applies equally well to noncommutative QFT's or to matrix models.
The key non-perturbative bound is indeed a norm bound: $H \ge 0$ implies $\Vert \frac{1}{1+H}) \Vert  \le 1$.
This is the exact analog of the  loop-vertex bound, in which an operator $\sigma$ Hermitian
implied  $\Vert \frac{1}{1+i \sigma}) \Vert  \le 1$ \cite{R1}-\cite{MR1}. These bounds extend exactly in the same way 
to matrix models. But the advantage of this formalism is that in the parametric version,
ordering the tree prepares exactly the theory for (continuous) multiscale analysis
and renormalization. Ultraviolet divergences in this formalism simply occur when
$D \ge 2$ in the form of divergence of certain matrix elements of the scalar product which define the quantum field theory.
Hence this scalar product itself should be defined inductively over scales
using running constants to absorb the divergences as usual \cite{Riv1}.

\section{The case of noninteger $D>0$}
\setcounter{equation}{0}
\label{nonint}

We would like to define QFT non perturbatively in non integer 
(positive) dimension.

The key should be  given by the parametric representation. It was remarked 
very early \cite{Speer} that Feynman amplitudes in parametric space only involve Gaussian
integrations and the result is therefore given in terms of determinants ("Symanzik polynomials") to the power $D$ times quadratic forms in the external invariants which are rotation invariant so involve only scalar external invariants. 

We start again from formula (\ref{parainteger}), but we want to perform all momentum and spatial integrations
to obtain determinants raised to the power  $D$. Let us rewrite $K$ in this form.
For each value of the functional derivatives  
\be  \prod_{i=1}^p   \frac{\delta}{\delta \phi(x_i)}   \biggl( \frac{\delta}{\delta \phi(x_0)}   \prod_{\ell \in T}   \frac{\delta}{\delta \phi(x_\ell)} \frac{\delta}{\delta \phi(x'_\ell)}   \biggr)
\ee we obtain a piece of a Feynman amplitude for a particular graph. Let us generically 
call $G$ a label for all these pieces of graphs. In any such $G$ the $x_i$ are now identified with
some particular $x_v$'s, and  we get a particular function $\chi_G (s, t) = \prod_{i=1}^p \chi ( s^T_i  \le t_i ) $.
It is then not difficult to perform for each such $G$ all internal spatial integrations $\prod_v d^D  x_v$ in 
$K$. The Feynman-Symanzik parametric representation then states that  $K$ 
is a sum of quadratic forms on the invariants built on the external variables which are now 
$z_T$, and the $ \{y\}$,
divided by a polynomial in the Feynman parameters, called $U_G (s, t)$ 
to the power $D/2$.

\be  \label{formofK} 
K_p (s_0, \{y\} ,  \{ t\}) = \sum_G c_G    \prod_{\ell \in T} 
\int_0^\infty d s_{\ell}  
\chi_G (s, t)  e^{Q_G ( \{y\} , z_T , s_0,   \{ t\}) } / U_G^{D/2} ( s, t ) 
\ee

The last remaining difficulty is that we still have to perform the integrals $d^D y_i$ in 
\bea  \label{integraly}   I =\sum_{p \ge 0}  \frac{1}{p!} \int_0^\infty ds_0    
\prod_{i=1}^p   \int_{s_0}^\infty dt_i \prod_{i=0}^p    \int  d^Dy_i 
\biggl(  K_p (s_0, \{y\} , \{ t\} )    \biggr)^2
\label{kernel2}
\eea
to get a formula in which the dimension purely enters as a parameter.
We can put $y_0$ to the origin to break translation invariance.
Substituting the form (\ref{formofK}) into  (\ref{integraly})
we can put the result into  the form
\be \label{quadr} \sum_{a,b}  \lambda_a \lambda_b  \det (Q_a + Q_b)^{-D/2}
\ee
where $Q_a$ is a $p$ by $p$ 
positive quadratic form on the invariants built on the $y_i$ variables,
and $a$ is a simpler label for the piece of graph $G$ under consideration.

We conjecture that such quadratic forms are always positive:

\medskip\noindent{\bf Conjecture 1}
{\it Let  $Q_a$, $a=1,..., q$ be a family of $q$ $p$ by $p$ positive quadratic forms with positive coefficients
Then the matrix $M_{ab} =  \det (Q_a + Q_b)^{-D/2} $ is positive for any $D>0$.}

\medskip
The conjecture is obviously true for integer $D$, and  for $q=2$ and any $D$ and $p$, since 
a two by two symmetric matrix with positive coefficients 
\be M = 
\begin{pmatrix} a & c \\ c & b  \end{pmatrix}
\ee
is positive if and only any of its Hadamard positive power 
\be M_d = \begin{pmatrix} a^d & c^d \\ c^d & b^d  \end{pmatrix}  
\ee is positive;
indeed positivity in that case reduces to the condition $ab \ge c^2$. However this is no longer true for 
$q\ge 3$. It is easy to check eg that for positive $r$ 
\be  M (r) = 
\begin{pmatrix} 1 & r & 0 \\  r & 1 & r \\ 0 &  r & 1  \end{pmatrix}
\ee
is positive if and only if $r\le 1/\sqrt {2}$; hence
\be  M (4/5) = 
\begin{pmatrix} 1 & .8 & 0 \\  .8 & 1 & .8 \\ 0 &  .8 & 1  \end{pmatrix}
\ee
is not positive but its Hadamard square
\be  M_2 = M(16/25) =
\begin{pmatrix} 1 & .64 & 0 \\  .64 & 1 & .64 \\ 0 & .64 & 1  \end{pmatrix}
\ee
is positive. We conjecture however that matrices obtained
from determinants of sums of positive quadratic forms as in (\ref{quadr})
are never of this kind.

The conjecture if true would 
lead to the first non perturbative definition of QFT's in non-integer dimension.

\section{Complex parameters, Borel Summability}
\setcounter{equation}{0}

It is natural to expect that for complex coupling constant
sufficiently small with positive real part, the formula 
(\ref{maindef2point}) still makes sense, and that  for $0< \Re D<2$
our non-perturbative definition is in fact
the Borel sum of the ordinary perturbative series.

\medskip\noindent{\bf Conjecture 2}
{\it The two point function $S_2$ is well defined 
by (\ref{maindef2point}) for $\lambda \phi^4_D$ for $0< \Re D<2$, and for
in a Nevanlinna-Sokal disk $\Re \lambda^{-1} \ge R^{-1}$. It is the 
Borel sum of its perturbative series.
}

\medskip
We do not expect the proof of this conjecture to be very difficult, but the resolvent
$H$ is not linear in $\lambda$ so the problem looks more like analyticity of a contnued fraction
rather than of a simple resolvent as was the case in \cite{R1}-\cite{MR1}.

If Conjecture 1 is true, we expect this Conjecture 2 to extend at least 
to the region $0\le D \le 4$, if we add the ultraviolet subtractions
corresponding to the mass renormalizations for $2 \le \Re D<2$.
The local "Borel germ" of perturbation theory was shown to exist 
in that region  $0\le D \le 4$ in \cite{SpeerRiv}.

The proof of Conjecture 2 involves presumably to define complex suitable extensions of
the real vector space ${\cal E}$ and the real symmetric $H$ operator. This and many other
applications of our formalism are devoted
to future publications.


\begin{thebibliography}{99} 


\bibitem{BLL}
F. Bergeron, G. Labelle and P. Leroux,
Combinatorial Species and Tree-like Structures (Encyclopedia of Mathematics and its Applications),
Cambridge University Press (1997).


\bibitem{BK} D. Brydges and T. Kennedy, 
Mayer expansions and the Hamilton-Jacobi equation,
Journal of Statistical Physics, {\bf 48}, 19 (1987).

\bibitem{AR1} A. Abdesselam and V.  Rivasseau, Trees, forests and jungles: a
botanical garden for cluster expansions, in Constructive Physics, ed by
V. Rivasseau, Lecture Notes in Physics 446, Springer Verlag, 1995, arXiv:hep-th/9409094.

\bibitem{HV} G. `t Hooft and M. Veltman, Regularization and 
Renormalization of gauge fields, Nucl. Phys. {\bf B44} No. 1, 189-213 (1972).

\bibitem{Wil}
K. Wilson, Quantum Field - Theory Models in Less Than 4 Dimensions,
Phys. Rev. D 7, 2911 - 2926 (1973).


\bibitem{A1} A. Abdesselam,
The Grassmann-Berezin calculus and theorems of the matrix-tree type". 
Adv. in Appl. Math. 33 (2004), no. 1, 51--70

\bibitem{GR} R. Gurau and V. Rivasseau, Parametric Representation of 
Noncommutative Field Theory, math-ph/0606030, Commun. Math. 
Phys. {\bf 272}, 811-835 (2007).

\bibitem{RT} V. Rivasseau and A. Tanasa,
Parametric representation of Critical noncommutative QFT models,
arXiv:math-ph/0701034, Commun. Math. Phys. Commun. Math. Phys.
{\bf 279}, 355-379, (2008).


\bibitem{Les} A. Lesniewski, Effective Action for the Yukawa$_{2}$ 
Quantum Field Theory, Commun. Math. Phys. {\bf 108}, 437 (1987).

\bibitem{FMRT1} J. Feldman, J. Magnen, V. Rivasseau and E. Trubowitz,
An Infinite Volume Expansion for Many Fermion Green's
functions, Helv. Phys. Acta, {\bf 65}, 679 (1992).

\bibitem{AR2} A. Abdesselam and V. Rivasseau, Explicit Fermionic Cluster
Expansion, Lett. Math. Phys. {\bf 44}, 77-88 (1998), 
arXiv:cond-mat/9712055.

\bibitem{DR1} M. Disertori and V. Rivasseau,
Continuous Constructive Fermionic Renormalization, 
Annales Henri Poincar{\'e}, {\bf 1}, 1 (2000), arXiv:hep-th/9802145. 

\bibitem{DR2} M. Disertori and V. Rivasseau, Interacting Fermi liquid 
in two dimensions at finite temperature, Part I: Convergent Attributions,
Commun. Math. Phys. {\bf 215}, 251 (2000); Part II: Renormalization,
in two dimensions at finite temperature, Part I: Convergent Attributions,
Commun. Math. Phys. {\bf 215}, 291 (2000).

\bibitem{FKT} Joel Feldman, Horst Kn\"orrer and Eugene Trubowitz,  
Commun. Math. Phys. 247 (2004): A Two Dimensional Fermi Liquid. Part 1: Overview,
1-47; Part 2: Convergence,  49-111; Part 3: The 
Fermi Surface, 113-177; Particle–Hole Ladders, 179-194;
Convergence of Perturbation Expansions in Fermionic Models.  Part 1: Nonperturbative Bounds, 195-242; Part 2: Overlapping Loops, 243-319.

\bibitem {Hub} V. Rivasseau, The two dimensional Hubbard Model at half-filling:
I. Convergent Contributions, Journ. Stat. Phys. Vol {\bf  106}, 693-722 (2002);
S. Afchain, J. Magnen and V. Rivasseau, 
Renormalization of the 2-point function of the Hubbard Model at half-filling, 
Ann. Henri Poincar\'e {\bf 6}, 399, (2005);
The Hubbard Model at half-filling, part III: the lower bound on the self-energy,
Ann. Henri Poincar\'e {\bf 6}, 449 (2005)

\bibitem{BGM} G. Benfatto, A. Giuliani and V. Mastropietro,
Low Temperature Analysis of Two-Dimensional Fermi Systems with Symmetric Fermi Surface,
Ann. Henri Poincar\'e {\bf 4} 137-193 (2003);
Fermi Liquid Behavior in the 2D Hubbard Model at Low Temperatures,
Ann. Henri Poincar\'e {\bf 7}, 809-898 (2006).

\bibitem{GW} H.~Grosse and R.~Wulkenhaar, ``Renormalization
of $\phi^4$-theory on noncommutative ${\mathbb R}^4$ in the matrix
base,'' Commun.\ Math.\ Phys. {\bf 256}, 305-374 (2005), arXiv:hep-th/0401128. 

\bibitem{GW2}
H.~Grosse and R.~Wulkenhaar, Power-counting theorem for non-local matrix
models and renormalization, {\em Commun. Math. Phys.} {\bf 254}, 91-127 (2005),
arXiv:hep-th/0305066.

\bibitem{RVW}
V.~Rivasseau, F.~Vignes-Tourneret, and R.~Wulkenhaar, Renormalization of
noncommutative $\phi^4$-theory by multi-scale analysis, {\em Commun. Math.
Phys.} {\bf 262}, 565--594 (2006),
arXiv:hep-th/0501036.

\bibitem{GMRV}
R.~Gurau, J.~Magnen, V.~Rivasseau and F.~Vignes-Tourneret, Renormalization
of non-commutative $\phi^4_4$ field theory in $x$ space, {\em
Commun.~Math.~Phys.} {\bf 267}, 515-542 (2006), 
arXiv:hep-th/0512271.

\bibitem{GrWubeta}
H.~Grosse and R.~Wulkenhaar, The beta-function in duality-covariant
noncommutative $\phi^{4}$-theory, {\em Eur. Phys. J.} {\bf C35} (2004),
277--282, hep-th/0402093.

\bibitem{DRbeta}
M.~Disertori and V.~Rivasseau, Two and Three Loops Beta Function of Non Commutative
$\Phi^4_4$ Theory, Eur. Phys. Journ. C {\bf 50} (2007), 661,
 hep-th/0610224.

\bibitem{DGMR}
M. Disertori, R.~Gurau, J.~Magnen and V.~Rivasseau,
Vanishing of Beta Function of Non Commutative $\Phi_4^4$ to all orders, 
Physics Letters B, 649 (1), p.95-102, (2007),  hep-th/0612251.

\bibitem{GJ} J. Glimm and A. Jaffe,
Quantum physics. A functional integral point of view, Springer, 2nd edition (1987).

\bibitem{Riv1} V. Rivasseau, From perturbative to constructive renormalization,
Princeton University Press (1991).

\bibitem {Br} D. Brydges, Weak perturbations of massless Gaussian measures, in 
Constructive Physics, LNP 446, Springer 1995.

\bibitem{AR3} 
A. Abdesselam and V. Rivasseau, An Explicit Large Versus Small Field Multiscale Cluster Expansion, 
Rev. Math. Phys. {\bf 9}, 123 (1997), arXiv:hep-th/9605094.

\bibitem{R1} V. Rivasseau, Constructive Matrix Theory, 
arXiv:hep-ph/0706.1224, JHEP {\bf 09} (2007) 008.

\bibitem{MR1} J. Magnen and V. Rivasseau, Constructive field theory without tears, 
arXiv:math/ph/0706.2457, to appear in Ann. Henri Poincar\'e.

\bibitem{Speer} E. Speer, Generalized Feynman Amplitudes,
Princeton University Press, 1969.

\bibitem{SpeerRiv}  E. Speer and V. Rivasseau,
The Borel transform in Euclidean $\phi^{4}_{4}$, Local existence 
for $Re\; \nu <4$, Commun. Math. Phys. {\bf 72}, 293 (1980).

\end{thebibliography}
\end{document}